\documentclass[11pt]{article}
\pdfoutput=1

\usepackage[T1]{fontenc}              
\usepackage[latin9]{inputenc}         

\usepackage[a4paper]{geometry}        

\usepackage{url}

\usepackage{booktabs}

\usepackage{graphicx} 
\usepackage{amssymb,amsfonts,amsmath}

\usepackage{natbib}
\usepackage{mathpazo}
\usepackage{microtype}

\usepackage{color}

\newcommand\MALDIquant{{\tt MALDIquant}\ }
\newcommand\readBrukerFlexData{{\tt readBrukerFlexData}\ }
\newcommand\readMzXmlData{{\tt readMzXmlData}\ }

\newcommand{\figcite}[1]{Fig.~\textbf{\ref{#1}}}

\begin{document}

\date{4 July 2012}

\title{
MALDIquant: a versatile R package for the analysis of mass spectrometry data
}

\author{Sebastian Gibb
      \thanks{Institute for Medical Informatics,
      Statistics and Epidemiology,
      University of Leipzig,
      H\"artelstr. 16--18,
      D-04107 Leipzig, Germany} 
{} and Korbinian Strimmer \footnotemark[1]
}

\maketitle
\begin{abstract}

\noindent\textbf{Summary:}
 {\tt MALDIquant} is an R package providing a complete
and modular analysis pipeline for quantitative analysis of mass spectrometry
data.  {\tt MALDIquant} is specifically designed with application 
in clinical diagnostics in mind and implements
 sophisticated routines for importing raw data,
preprocessing, non-linear peak alignment, and calibration. It also 
handles technical replicates as well as spectra with unequal resolution.

\noindent\textbf{Availability:}
{\tt MALDIquant} 
and its associated R packages \\
{\tt readBrukerFlexData} and
{\tt readMzXmlData} are 
freely available
from the R archive CRAN (\url{http://cran.r-project.org}).
The software is distributed under the GNU General Public License
(version 3 or later) and is accompanied by
example files and data. Additional documentation
is available from \url{http://strimmerlab.org/software/maldiquant/}. 

\noindent\textbf{Contact:} \url{mail@sebastiangibb.de}

\end{abstract}

\newpage

\section{Introduction}

Mass spectrometry profiling is increasingly becoming an important tool in clinical diagnostics, for example to identify biomarkers for cancer \citep[e.g.][]{FL+2009}. Similarly as with other high-throughput technologies, 
sophisticated statistical algorithms are essential in the analysis
of spectrometry data \citep{MB+2010}. 

We have developed \MALDIquant to provide a
complete open source analysis pipeline on the R platform \citep{RPROJECT}
comprising all steps from importing of raw data, preprocessing 
(e.g. baseline removal), peak detection,  non-linear peak alignment
to calibration of mass spectra.
\MALDIquant is written as a standalone package 
using S4 object oriented programming to facilitate further extension.

\MALDIquant was initially developed for clinical proteomics 
using MALDI (Matrix-Assisted Laser Desorption/Ionization) technology.
However, the algorithms implemented in \MALDIquant are generic and 
may be equally applied to other 2D mass spectrometry data.

\section{Distinctive Features}

In comparison with related R packages for mass spectrometry analysis
\MALDIquant features a number of unique capabilities. In particular, it
implements a sophisticated non-linear peak alignment algorithm
\citep{WF+2010,HW+2011} as well as a calibration procedure for
normalization of peak intensities across spectra 
that is modeled on a related method for sequence 
count data \citep{AH2010}.
In addition, \MALDIquant allows to analyze technical replicates and 
spectra with unequal resolution, a crucial feature in
 clinical mass spectrometry where spectra from multiple
sources need to be compared. 

\section{Details on Algorithms}

\begin{figure*}[!ht]
\centering
\includegraphics[width=\textwidth]{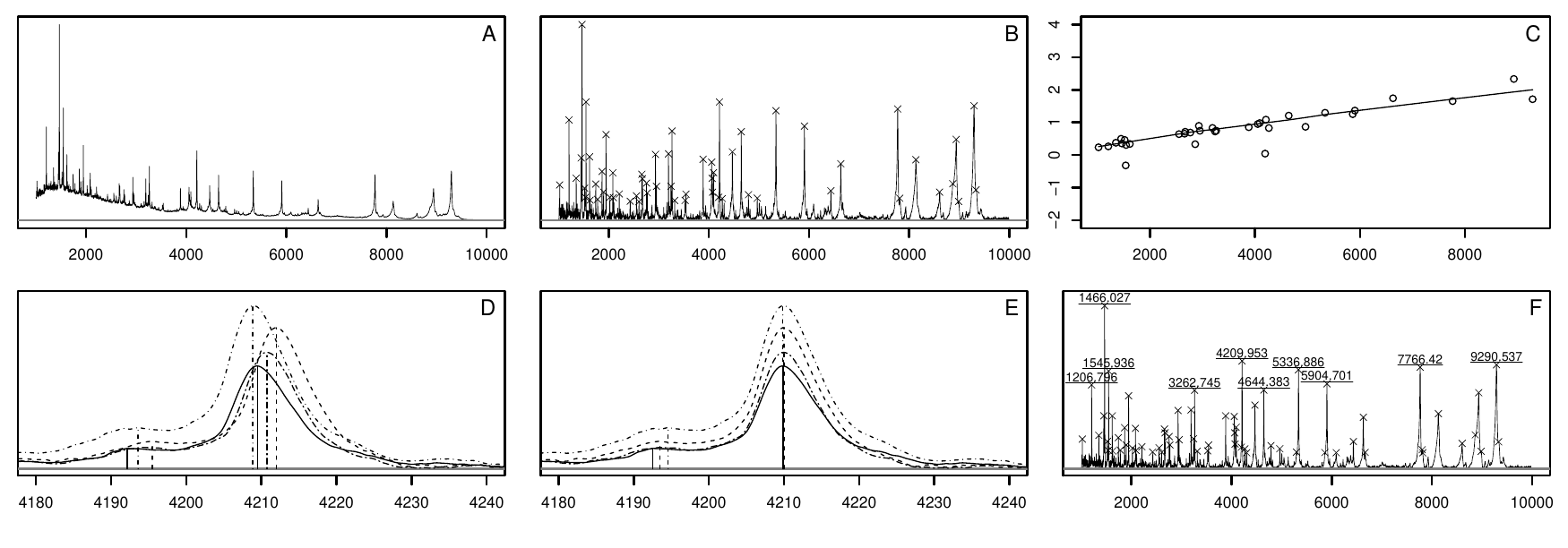}
\vspace{-1cm}
\caption{Example of \MALDIquant output: A: raw spectrum;  
B: variance-stabilized, smoothed, baseline-corrected spectrum with detected peaks; 
C: fitted warping function for peak alignment; 
D: four unaligned peaks;
E: four aligned peaks; 
F:  merged spectrum with detected and labeled peaks.
}
\label{fig:one}
\end{figure*}

An example workflow for mass spectrometry analysis using
\MALDIquant is depicted
in \figcite{fig:one}, starting with a raw unprocessed MALDI spectrum (A),
followed by smoothing, baseline correction and peak detection (B), local alignment of
peaks across spectra by warping (C--E), and merging and visualization (F).
In the following we briefly provide some background on the respective algorithms.

\subsection{Data import}

\MALDIquant is carefully  designed
to be independent of any specific mass spectrometry hardware. 
Nonetheless, native input of binary data files
(as well as complete folder hierarchies)  from Bruker *flex series 
instruments and input of the mzXML data format is supported
via the associated R packages \readBrukerFlexData and \readMzXmlData.

\subsection{Data preprocessing}

For preprocessing spectral data \MALDIquant offers a complete set of
routines for
smoothing, variance stabilization, baseline correction, and peak detection.
\MALDIquant implements several approaches to adjust the baseline, and uses
per default
the SNIP algorithm \citep{RC+1988} that returns a smooth baseline 
and leads to positive corrected intensities (\figcite{fig:one}B).

\subsection{Peak alignment}

For comparison of peaks across different spectra it is essential to 
conduct alignment. In order to match peaks belonging to
the same mass  \MALDIquant uses a statistical regression-based approach combining 
the algorithms of   \citet{WF+2010} and \citet{HW+2011}.
Specifically, first landmark peaks are identified that occur in most
spectra.  Subsequently, a non-linear warping function is computed for
each spectrum by fitting a local regression to the matched reference peaks (\figcite{fig:one}C--E).   This also allows to merge 
 aligned spectra from technical replicates.  An example of a merged spectrum
with identified and labeled peaks is shown in  \figcite{fig:one}F.

\subsection{Calibration}

Quantitative analysis of multiple spectra, e.g. to detect differentially
expressed peaks, requires calibration.  In order to render peak intensities comparable
across spectra a suitable scale factor for each individual
spectrum needs to be determined.  Experimentally, quantification of intensities is performed by reference to spike-in samples. In absence of spike-ins \MALDIquant 
offers a way of calibrating relative intensities by adapting 
an algorithm for calibrating
next generation sequencing data \citep{AH2010}. In this procedure 
first a reference spectrum is created using the median intensity 
of aligned peaks from all spectra.
Subsequently, a 
scale factor is computed for each spectrum by employing a robust 
estimator of the overall ratio of the peak intensities of the uncalibrated spectrum versus the reference
spectrum. Additionally, calibration based on
 total ion current (TIC) is available.

\subsection{Classification and feature selection}

Finally, the resulting  calibrated peak intensity matrix  may be exported for further use in high-level statistical
analysis, for instance classification and feature selection using
shrinkage discriminant analysis \citep{AS2010}.

\section{Conclusion}
\MALDIquant is a versatile R package providing a flexible analysis pipeline for
MALDI-TOF and other mass spectrometry data.   It offers a number of 
distinctive features, in particular for alignment by non-linear warping and simultaneous calibration of peak intensities.  
 
An overview of  its capabilities is given by running the included
demo script
\begin{verbatim}
    library("MALDIquant")
    demo("MALDIquant")
\end{verbatim}

\section*{Acknowledgments}

We thank Alexander Leichtle 
for many valuable and helpful 
suggestions and \citet{FL+2009}
for their kind permission to use their data in {\tt MALDIquant}.


\bibliographystyle{apalike}
\bibliography{preamble,stats,med,array,ngs}

\end{document}